\documentclass[aoas]{imsart}

\RequirePackage{amsthm,amsmath,amsfonts,amssymb}
\RequirePackage[colorlinks,citecolor=blue,urlcolor=blue]{hyperref}
\RequirePackage{graphicx}
\RequirePackage[authoryear]{natbib}
\usepackage{booktabs}
\usepackage{float}
\usepackage{comment}

\startlocaldefs
\theoremstyle{plain}

\newtheorem{theorem}{Theorem}[section]
\newtheorem{lemma}[theorem]{Lemma}
\newtheorem{corollary}[theorem]{Corollary}
\theoremstyle{remark}

\newtheorem{proposition}[theorem]{Proposition}

\theoremstyle{remark}
\newtheorem{remark}{Remark}

\newcommand{\indep}{\perp \!\!\! \perp}

\renewcommand{\qed}{\hfill\rule{2mm}{2mm}}
\newcommand\E{\mathbb{E}}
\newcommand{\PP}{\mathbb{P}}
\newcommand{\norm}[1]{\left\lVert #1 \right\rVert}

\endlocaldefs

\begin{document}
\begin{frontmatter}
    \title{Effects of adolescent victimization on offending: flexible methods for missing data \& unmeasured confounding}

\begin{aug}
\author[A,B]{\fnms{Mateo}~\snm{Dulce Rubio}\ead[label=e1]{mdulceru@andrew.cmu.edu}},
\author[A]{\fnms{Edward H.}~\snm{Kennedy}}
\\ \author[C]{\fnms{Valerio}~\snm{Baćak}}
\and
\author[B]{\fnms{Daniel S.}~\snm{Nagin}}
\address[A]{Department of Statistics and Data Science, Carnegie Mellon University\printead[presep={\\ }]{e1}}

\address[B]{Heinz College of Information Systems and Public Policy, Carnegie Mellon University}

\address[C]{School of Criminal Justice, Rutgers University}
\end{aug}

\begin{abstract}

    The causal link between victimization and violence later in life is largely accepted but has been understudied for victimized adolescents. In this work we use the Add Health dataset, the largest nationally representative longitudinal survey of adolescents, to estimate the relationship between victimization and future offending in this population. To accomplish this, we derive a new doubly robust estimator for the average treatment effect on the treated (ATT) when the treatment and outcome are not always observed. We then find that the offending rate among victimized individuals would have been 3.86 percentage points lower if none of them had been victimized (95\% CI: [0.28, 7.45]). This contributes positive evidence of a causal effect of victimization on future offending among adolescents. We further present statistical evidence of heterogeneous effects by age, under which the ATT decreases according to the age at which victimization is experienced. We then devise a novel risk-ratio-based sensitivity analysis and conclude that our results are robust to modest unmeasured confounding. Finally, we show that the found effect is mainly driven by non-violent offending.

\end{abstract}


\end{frontmatter}

\section{Introduction}

There is a common belief that violence begets violence. This perception is often grounded in the \textit{Cycle of Violence} theory which asserts that victimized children are more likely to engage in violent behavior in the future (\cite{cov_widom89}). Such theory has been the foundation of several public policies in recent decades that have sought to mitigate this association and protect maltreated individuals (\cite{cov_update_widom01}). However, this relationship has been scarcely studied among adolescents, who suffer high rates of victimization. It is thus crucial for policymakers to identify and understand whether such a causal relationship exists for this population.

The cycle of violence theory has been empirically supported in the literature. For instance, in their seminal work, \cite{cov_widom89, cov_update_widom01, intergenerational_widom15} found that maltreated infants exhibit violent or delinquent behaviors more frequently than those not abused, using a longitudinal study with 908 individuals abused or neglected during childhood and a matched control group of 667 subjects not recorded as abused or neglected. The treatment and control groups were matched by sex, age, race, and socioeconomic status. These results have been replicated in recent studies using other datasets and diverse definitions of maltreatment and offending, additional control variables, and different methodological designs (\cite{understanding_currie12, revisiting_myers18, cov_context_wright13}). Such studies provide empirical evidence supporting the cycle of violence theory by analyzing the effect of different types of child victimization (physical abuse, sexual abuse, and neglect) on future violent behavior. In general, the literature seems to agree that child abuse does have an unfavorable causal effect on future misbehavior.

However, there are some theories that claim that this effect varies over time, such as the developmental psychopathology and the life course theories (\cite{unsafe_mersky}). In particular, adolescents have different cognitive development and thus potentially distinct consequences of experiencing victimization. Moreover, this is of particular importance given that, according to the Office of Juvenile Justice and Delinquency Prevention, adolescents and young adults face victimization rates almost twice as high as those of people over the age of 25 (\cite{ojjdp_rates}). Despite the high victimization rate, its relationship with future offending is poorly understood.

For instance, \cite{unsafe_mersky} compares the effects of childhood and adolescent victimization on future delinquency and crime. The authors study whether the effect on future offending varies with the age at which victimization is experienced, concluding that the timing of victimization does influence this effect. The authors show that childhood maltreatment is consistently positively associated with violent behavior later in life, while the relationship between adolescent victimization and adult crime, although positive, is less robust. 

A closely related question has been discussed in the literature on bullying, studying adolescent victimization perpetrated by a peer with a perceived power advantage, and potentially repeated over time (\cite{ojjdp_bullying, olweus_def}). In this regard, \cite{ttofi_as_predictor} conducts a systematic review of the literature studying the relationship between school bullying victimization and future violent behaviors. Based on 12 relevant articles, the authors conclude that being a victim of school bullying increases the likelihood of being violent by one-third, even after controlling for other risk factors. 

\section*{Methodological Challenges} Despite the importance and abundance of adolescent victimization, understanding the consequences of this phenomenon imposes non-negligible methodological difficulties. First, to identify the causal effect of interest there are some assumptions that need to be made. In particular, there should not be confounding factors that make victimized individuals systematically different from those not victimized in the control group, which in turn affect future violent behavior (e.g., socioeconomic status) (\cite{imbens_rubin}). Additional biases can be introduced by the use of often implausible parametric assumptions, or there can be a loss of efficiency when using simple but typically sub-optimal `plug-in' estimators (\cite{ehk_semiparametric}).

Furthermore, the literature has identified what has been called the \textit{Victim-Offender Overlap} in which victimized individuals are also victimizers (\cite{victim_offender_overlap}). Such overlap makes it difficult to identify whether victimization is the cause or the effect of violent behavior (\cite{offending_on_victimization}). To overcome this problem, it is necessary to use panel data to ensure that the victimizing event actually occurred before the criminal behavior to be analyzed. However, panel data pose an additional challenge if there is attrition not completely at random, especially if such attrition makes the treatment or the outcome not always observable (\cite{missing_exposures, zhang2016causal}).

\section*{Contributions}
In this work, 
we use the Add Health longitudinal dataset and flexible modeling to study the relationship between victimization and future offending among adolescents. We derive a novel nonparametric doubly robust estimator of the average treatment effect on the treated (ATT) in the presence of missing data, and show it is root-n-consistent, asymptotically normal, and optimally efficient in a flexible nonparametric model. Our methods account for possible confounding factors associated with both treatment (victimization) and outcome (offending), and handle the missing information due to attrition, which makes the treatment and outcome partially missing. We find a positive effect of victimization on future offending among adolescents: the offending rate would have been 33.5\% lower (95\% CI: [2.4\%, 64.7\%]) if none of the victimized adolescents had been victimized.

Moreover, we estimate heterogeneous effects by age and use a doubly robust hypothesis test to conclude that the effect of victimization on future offending varies depending on the age at which the victimizing event is experienced. We then derive a new risk-ratio-based sensitivity analysis for the no unmeasured confounding assumption and find that our results are robust to modest confounding. Finally, we note that our results are mainly driven by future engagement in non-violent crimes.

\section*{Paper Outline} Section \ref{sec:data} describes the Add Health dataset we use to study the relationship between victimization and future offending among adolescents. In Section \ref{sec:methods} we detail the proposed methodology, state our causal assumptions and derive a novel doubly robust estimator for the ATT. Section \ref{sec:results} presents our main results and analyzes the robustness of our findings. Finally, Section \ref{sec:discussion} concludes and discusses the limitations of our work. All the proofs are detailed in the Appendix.
\section{Data}
\label{sec:data}

In this section, we describe the dataset and data structure used to study the effect of victimization on future offending among adolescents. The dataset chosen and the data structure considered are intended to address the methodological challenges described above: the victim-offender overlap and missingness not completely at random.


We use the National Longitudinal Study of Adolescent to Adult Health dataset (Add Health). Add Health is the largest longitudinal survey of adolescents in the United States (\cite{addhealth}). The study followed a nationally representative sample of students in grades 7-12 during the 1994-95 school year (between 11 and 21 years old), with surveys conducted in 1995, 1996, 2001-02, 2008, and 2016-18. In addition, the study administered questionnaires to parents, siblings, fellow students, school administrators, and romantic partners. Other datasets provide information about the neighborhoods and communities of the respondents (\cite{addhealth}). Such a rich dataset allows us to incorporate diverse confounding factors in our analysis.

In this work, we use the public-use dataset\footnote{\url{https://addhealth.cpc.unc.edu/data/\#public-use}} which includes a random sample of 6,504 respondents surveyed during Wave 1 (1994-95), Wave 2 (1996), and Wave 3 (2001-02). We exploit the temporality of the study to ensure that the control variables are measured prior to the treatment (adolescent victimization), which in turn occurs before the outcome of interest (future offending). 

More formally, we observe the following data for each respondent
$$(X, R, RA, RY),$$
where $X$ is a set of control variables measured at Wave 1 (in 1995), $A$ is the treatment variable of being victimized measured at Wave 2 (in 1996), $Y$ is the outcome of interest of offending at Wave 3 (in 2001), and $R$ is an indicator variable of whether an individual was observed at all the three waves.

We define the treatment $A$ for each respondent (Victimization at Wave 2) as whether in the 12 months prior to the survey \textit{someone pulled a knife or gun at them, someone shot them, someone stabbed them}, or \textit{they were beaten}. Similarly, we define the outcome $Y$ (Offending at Wave 3) as whether in the 12 months prior to the survey the respondent \textit{pointed a knife or gun at someone}, \textit{shot or stabbed someone}, \textit{stole something worth more than \$50}, \textit{broke into a house or building to steal something (burglary)}, or \textit{used or threatened to use a weapon to get something from someone (robbery)}. It should be noted that this information is self-reported by the respondents.
 
To account for potential confounders we include in $X$ the following socioeconomic and demographic variables collected at Wave 1: \textit{Sex, Race, Born in the US, Age, Working, Presently in school, Marital status, Repeated a grade,  Perceived importance of post-HS education, Expectation of post-HS education, Self-rated health, Should get medical care but did not, Considered suicide, Parental supervision, Got drunk more than once a month, Illicit drug use, Self-control, People care in neighborhood, Feel safe in neighborhood, Parents receive welfare, Parents have enough money, Victimization at Wave 1,} and \textit{Offending at Wave 1}. The Appendix reports descriptive statistics for these control variables. Note that we include as controls whether the individuals were victimized or committed an offense in the 12 months before Wave 1 to control for these baseline levels. 

Finally, due to attrition, 26.1\% of the sample has missing information on the treatment (victimization at Wave 2), and 25.9\% lacks information on the outcome (offending at Wave 3). This missingness is likely not completely at random since, for example, being victimized may affect the likelihood of being observed thereafter. Ignoring the missingness would thus likely lead to inaccurate causal conclusions; therefore, we employ methods that allow the missingness to depend on covariate information, as detailed in the next section.

\section*{Notation}

We write $\PP_n(Z) =\frac{1}{n}\sum_{i=1}^{n} Z_i$ as the sample average of any $Z_i$, $i=1,...,n$. We write $T_n=o_{\PP}(r_n)$ and $T_n=O_\PP(r_n)$ to mean that $T_n/r_n$ converges in probability to zero and is bounded in probability, respectively. As shorthand we define 
\begin{align*}
\omega(X) &= \PP[R=1|X] \\
\pi(X) &= \E[A|X, R=1] \\
\mu_0(X) &= \E[Y|X,R=1, A=0] \\
\mu_1(X) &= \E[AY|X,R=1] 
\end{align*}
as the missingness propensity score, treatment propensity score among the observed, and outcome regressions under control and treatment among the observed.

\section{Methods}
\label{sec:methods} 

In this section, we detail the methodology used to answer the research question of interest: the causal effect of victimization on future offending among adolescents. We motivate and derive a new doubly robust estimator for the average treatment effect on the treated (ATT) when the treatment and outcome are not always observed due to attrition. We also make explicit the assumptions required to draw causal conclusions in our analysis.

\section*{Causal Identification} We are interested in answering the causal question: \textit{how would the offending rate among victimized individuals have changed, if none of them had been victimized?} That is, we are interested in estimating the ATT parameter, given by
\begin{equation}
    \psi_{ATT} = \E\left[Y^1 - Y^0 \mid A=1\right].
\end{equation}
Here $Y^1$ denotes the potential offending outcome if the individual had been victimized, and $Y^0$ denotes the potential outcome if they had not been victimized. However, these variables are not fully observed, since each individual is either victimized or not, non-randomly. Thus, to identify the expected value $\E\left[Y^1 - Y^0 \mid A=1\right]$ using the observed data, we use the following causal assumptions:
\begin{itemize}
    \item[\textbf{(A1)}] \textbf{Consistency:} $Y = Y^a$ if $A=a$.
    \item[\textbf{(A2)}] \textbf{Positivity:} $\PP(A = 0 ~ \mid ~ X) \geq \epsilon > 0$ with probability 1.
    \item[\textbf{(A3)}] \textbf{No Unmeasured Confounders (NUC):} $A \indep Y^0 ~ \mid ~ X$.
\end{itemize}

The Consistency assumption (A1) states that a subject's outcome is uniquely determined by its own treatment level, and is not affected by the treatment of other units. Positivity (A2) assumes that everyone in our population of interest has some positive chance of not being victimized, given covariates. Given our target estimand, (A2) is weaker than the usual positivity assumption, which would also requires that everyone has a positive probability of being victimized. The third NUC assumption (A3) says that within covariate levels, the treatment assignment is as good as random. These are standard causal assumptions that have been extensively discussed in the literature (\cite{imbens2004nonparametric, van2003unified}). However, we are aware that these assumptions may be violated. Therefore, in the Results Section, we perform a sensitivity analysis relaxing the NUC assumption.

An added complication with the Add Health data is that, in addition to the usual problem of not observing both potential outcomes $Y^1$ and $Y^0$, we also do not always observe the treatment and outcome, due to attrition. To address this, we use the following assumptions regarding missingness:
\begin{itemize}
    \item[\textbf{(R1)}] \textbf{Positivity of follow-up:} $\PP(R = 1 ~ \mid ~ X) \geq \epsilon > 0$ with probability 1.
    \item[\textbf{(R2)}] \textbf{Missing At Random (MAR):} $R \indep (A, Y) ~ \mid ~ X$.
\end{itemize}
The Positivity of follow-up assumption (R1) says that each subject has some positive chance of being observed, conditioned on its covariates. The MAR assumption (R2) allows us to treat the missingness as good as random within covariate levels (\cite{williamson2012doubly, chaudhuri2016gmm, zhang2016causal}). Under these causal assumptions, we are able to identify the causal parameter of interest $\psi_{ATT}$ with the observed data provided by Add Health. 

\begin{proposition}
\label{prop:att_identification}
    \textbf{(Identification result)} Under assumptions (A1)-(A3) and (R1)-(R2), the ATT $\psi_{ATT}$ is identified as

    \begin{equation}
        \psi_{ATT} = \frac{\E[\E[AY \mid X, R=1]] - \E[\E[A \mid X,R=1]\E[Y \mid X,R=1,A=0]]}{\E[\E[A \mid X, R=1]]}.
    \end{equation}

\end{proposition}

Proofs of all results in the paper are given in Section D of the Appendix.

\section*{Statistical Estimation} The next step is to flexibly and efficiently estimate our parameter of interest $\psi_{ATT}$. For this task, there are a handful of options with different desirable properties. Among these, one-step or doubly robust estimators coming from semiparametric theory can be root-n consistent, asymptotically normal, and optimally efficient under mild assumptions \citep{ehk_semiparametric, van2003unified, semiparametric}. 

To this end, we use one-step estimators for each of the terms in the identifying expression given by Proposition \ref{prop:att_identification} and prove that the corresponding one-step estimator for $\psi_{ATT}$ has similar properties. In this regard, we make use of well-studied one-step estimators for the terms $\psi_{ay^1} = \E[\E[AY \mid X, R=1]]$ and $\psi_{a} = \E[\E[A \mid X, R=1]]$. These can be seen as standard parameters such as an Average Treatment Effect with the `treatment' being $R=1$ and the outcome being $AY$ and $A$, respectively. For a detailed derivation of such one-step estimators, we refer the reader to \cite{hines2022demystifying} or Example 2 throughout \cite{ehk_semiparametric}.

One-step estimators can be analyzed in a generic way since they are a sample average of an estimated function. As shown by \cite{ehk_semiparametric}, for a one-step influence function-based estimator $\widehat{\psi}$ for a functional $\psi$, we can decompose $\widehat{\psi} - \psi$ into three terms. The first term is a sample average of a fixed function and converges to a normal distribution after rescaling, by the Central Limit Theorem. The second term is an empirical process term that is of order $o_{\PP}(1/\sqrt{n})$ under a mild consistency assumption, and if, for example, appropriate sample splitting is used to construct $\widehat{\psi}$ (see Remark \ref{remark:ss}). The third term generally involves, for one-step estimators, second-order products of errors that can be negligible under relatively weak nonparametric conditions.

The one-step estimators for the three components of the ATT 
\begin{align*}
\psi_{ay^1} &= \E[\E[AY \mid X, R=1]] \\
\psi_a &= \E[\E[A \mid X, R=1]] \\
\psi_{ay^0} &= \E[\E[A \mid X,R=1]\E[Y \mid X,R=1,A=0]]
\end{align*}
are given by 
\begin{equation}
\label{eq:onestep_comp}
\widehat{\psi}_{ay^1}=\PP_n(\widehat\varphi_{ay^1}) \ , \ \ \widehat{\psi}_{a}=\PP_n(\widehat\varphi_{a}) \ , \ \ \widehat\psi_{ay^0}=\PP_n(\widehat\varphi_{ay^0}) \ , 
\end{equation}
where $(\widehat\varphi_{ay^1}, \widehat\varphi_{a}, \widehat\varphi_{ay^0})$ are estimates of the (uncentered) influence functions
\begin{align*}
\varphi_{ay^1} &= \frac{R}{{\omega}(X)}\left\{AY-{\mu}_1(X)\right\} + {\mu}_1(X) \\
\varphi_{a} &= \frac{R}{{\omega}(X)}\left\{A-{\pi}(X)\right\} + {\pi}(X)\\
\varphi_{ay^0} &= \frac{R}{{\omega}(X)}(A-{\pi}(X)){\mu}_0(X) + \frac{R}{{\omega}(X)}\frac{(1-A){\pi}(X)}{1-{\pi}(X)}(Y-{\mu}_0(X)) + {\pi}(X){\mu}_0(X)
\end{align*}
obtained by replacing the unknown nuisance functions $(\omega,\pi,\mu_0,\mu_1)$ 
with estimates. As noted above, the first two estimators  $\widehat{\psi}_{ay^1}$ and $\widehat{\psi}_{a}$ are by now standard doubly robust ATE estimators, while the third estimator $\widehat{\psi}_{ay^0}$ is new, to the best of our knowledge. In the following Lemma we analyze these estimators, on the way to analyzing the combined ATT estimator given by 
\begin{equation}
    \widehat{\psi}_{ATT} = \frac{\widehat{\psi}_{ay^1} - \widehat{\psi}_{ay^0}}{\widehat{\psi}_{a}}.
    \label{eq:onestep_att}
\end{equation}
In particular we give conditions under which they are root-n consistent and asymptotically normal, which  will then imply the root-n consistency and asymptotic normality of $\widehat\psi_{ATT}$.

\begin{remark}
\label{remark:ss}

   We use $K$-fold cross-fitting/sample-splitting to control the empirical process terms in our one-step estimators instead of introducing potentially restrictive assumptions on the class of functions $\{\varphi\}_{P \in \mathcal{P}}$ (e.g., Donsker), \citep{chernozhukov2018double, ehk_semiparametric}. Concretely, we use separate independent samples to estimate the components of the influence functions $\widehat{\varphi}_{ay^1}$, $\widehat{\varphi}_a$ and $\widehat{\varphi}_{ay^0}$, and to compute their empirical average $\widehat{\psi}_{ay^1}$, $\widehat{\psi}_{a}$, and $\widehat{\psi}_{ay^0}$, respectively. We then swap the samples to recover full efficiency by cross-fitting. A complete discussion on sample-splitting and alternative assumptions can be found in \cite{ehk_semiparametric}.
\end{remark}

\begin{lemma}
\label{lema:att_reminders}
Use $K$-fold cross-fitting with finite $K$ and assume:
\begin{enumerate}
\item $\norm{\widehat{\varphi}_{ay^1} - \varphi_{ay^1}} = o_{\PP}(1)$,
\item $\norm{\widehat{\varphi}_{a} - \varphi_a} = o_{\PP}(1)$, 
\item $\norm{\widehat{\varphi}_{ay^0} - \varphi_{ay^0}} = o_{\PP}(1)$, and 
\item $\PP(\widehat{\pi} \leq 1-\epsilon) = \PP(\epsilon \leq \widehat{\omega} \leq 1-\epsilon) = 1$ for some $\epsilon>0$.
\end{enumerate}
Then the proposed estimators satisfy
\begin{align*}
\widehat{\psi}_{ay^1} - \psi_{ay^1} &= (\PP_n - \PP)\left(\varphi_{ay^1} \right) + o_{\PP}(1/\sqrt{n}) + R_{2,ay^1} \\
\widehat{\psi}_{a} - \psi_a &= (\PP_n - \PP)\left(\varphi_a \right) + o_{\PP}(1/\sqrt{n}) + R_{2,a} \\
\widehat{\psi}_{ay^0} - \psi_{ay^0} &= (\PP_n - \PP)\left(\varphi_{ay^0} \right) + o_{\PP}(1/\sqrt{n}) + R_{2,ay^0}
\end{align*}
with the remainder terms bounded as
\begin{align*}
R_{2,ay^1} &\lesssim \norm{\widehat{\omega} - \omega}\norm{\widehat{\mu}_1 - \mu_1} \\
R_{2,a} &\lesssim \norm{\widehat{\omega} - \omega}\norm{\widehat{\pi} - \pi} \\
R_{2,ay^0} &\lesssim \norm{\widehat{\pi} - \pi}\norm{\widehat{\mu}_0 - \mu_0} + \norm{\widehat{\omega} - \omega}(\norm{\widehat{\pi} - \pi} + \norm{\widehat{\mu}_0 - \mu_0}).
\end{align*}
\end{lemma}

The previous result shows, in particular, that the one-step estimator $\widehat{\psi}_{ay^0}$ is root-n consistent, asymptotically normal, and optimally efficient if any two of the three nuisance functions $\pi, \omega, \mu_0$ are estimated consistently. Putting the pieces together, the following theorem analyzes the new one-step estimator for $\psi_{ATT}$ given in \eqref{eq:onestep_att}, which is one of  our main methodological contributions.

\begin{theorem}
\label{thm:onestep_att}
(\textbf{One-step estimator for $\psi_{ATT}$}) Use $K$-fold cross-fitting with finite $K$ and assume:
\begin{enumerate}
    \item $\norm{\widehat{\varphi}_{i} - \varphi_i} = o_{\PP}(1)$ for $i \in \{a, ay^1, ay^0\}$,
    \item $\PP(\epsilon < \widehat{\pi} < 1-\epsilon) = \PP(\epsilon < \widehat{\omega} < 1-\epsilon) = 1$.
    \item $\widehat{\psi}_{a} - \psi_{a} = o_{\PP}(1)$,
    \item $\PP(\epsilon < \widehat{\psi}_{a} < 1-\epsilon) = \PP(\epsilon < \psi_{a} < 1-\epsilon) = 1$.
\end{enumerate}
Then, the one-step estimator for $\psi_{ATT}$ given in \eqref{eq:onestep_att}, satisfies
$$\widehat{\psi}_{ATT} - \psi_{ATT} = (\PP_n - \PP)\left\{\varphi_{ATT} \right\} + o_{\PP}(1/\sqrt{n}) + R_{2,ATT},$$
where $\varphi_{ATT} = \frac{1}{\psi_{a}}\left\{\varphi_{ay^1} - \varphi_{ay^0} - \varphi_{a}\psi_{ATT} \right\}$ is the efficient influence function for $\psi_{ATT}$ and 
$$R_{2,ATT} \lesssim \norm{\widehat{\omega} - \omega}(\norm{\widehat{\pi} - \pi} + \norm{\widehat{\mu}_0 - \mu_0} + \norm{\widehat{\mu}_1 - \mu_1}) + \norm{\widehat{\pi} - \pi}\norm{\widehat{\mu}_0 - \mu_0}.$$

\end{theorem}

Theorem \ref{thm:onestep_att} implies that $\widehat{\psi}_{ATT}$ is a consistent estimator for $\psi_{ATT}$ if
\begin{itemize}
    \item $\omega$ and $(\pi$ or $\mu_0)$ are estimated consistently, or
    \item $\pi, \mu_0$ and $\mu_1$ are estimated consistently.     
\end{itemize}

To the best of our knowledge, this is the first one-step estimator derived for the ATT when the treatment and outcome are not always observed. Furthermore, the following two corollaries show the asymptotic normality of our estimator $\widehat{\psi}_{ATT}$ and give a recipe for computing confidence intervals easily.

\begin{corollary}   
\label{cor:normal_att}
    Under the assumptions of Theorem \ref{thm:onestep_att} and if $R_{2,ATT} = o_{\PP}(1/\sqrt{n})$, we have
    \begin{equation}
        \sqrt{n}\left(\widehat{\psi}_{ATT} - \psi_{ATT}\right) \rightsquigarrow \mathcal{N}\left(0, ~\sigma^2_{ATT} \right),
    \end{equation}
    where $\sigma^2_{ATT} = Var\left(\varphi_{ATT}\right) = Var\left(\dfrac{1}{\psi_{a}}\left\{\varphi_{ay^1}-\varphi_{ay^0} - \varphi_{a} \psi_{ATT}\right\}\right).$
\end{corollary}

\begin{corollary}   
     A $95\%$-level confidence interval for $\widehat{\psi}_{ATT}$ can be computed as 
\begin{equation}
\label{eq:ci_dr}
    \widehat{\psi}_{ATT} \pm 1.96 \sqrt{\frac{\widehat{\sigma}^2_{ATT}}{n}}.
\end{equation}
\end{corollary}

Summarizing, we have given methods and statistical guarantees for identifying and estimating the ATT when the treatment and outcome are partially missing. Our parameter of interest $\psi_{ATT}$ is identified under the standard causal assumptions of consistency, positivity, no unmeasured confounders, positivity of follow-up and missing at random, by the identifying expression in Proposition \ref{prop:att_identification}. Then, Theorem \ref{thm:onestep_att} says how to estimate $\psi_{ATT}$ using nonparametric theory building on previous results and a novel one-step estimator. Our proposed estimator is shown to be consistent, asymptotically normal, and optimally efficient in a flexible nonparametric model.

\section*{Sensitivity Analysis}

Following the approach proposed in \cite{stats_sensitivity}, we relax the NUC assumption and replace it with the assumption that the ratio between the mean potential outcome of treated individuals, had they not been treated, and the mean outcome of those who were not treated is bounded by a parameter $\delta > 1$:
\begin{equation}
\label{eq:sensitivity_ratio}
    \frac{1}{\delta} \leq \frac{\E[Y^0 | X, A=1, R=1]}{\E[Y^0 | X, A=0, R=1]} \leq \delta.
\end{equation}
Note that under the NUC assumption $Y^0 \indep A ~|~ X$, this ratio is equal to $1$. Thus, equation (\ref{eq:sensitivity_ratio}) is indeed a relaxation of the NUC assumption.

\begin{proposition}
\label{prop:att_sens}

Under assumptions (A1)-(A2), (R1)-(R2) and (\ref{eq:sensitivity_ratio}), we have that the causal parameter of interest $\psi_{ATT}$ satisfies
\begin{equation}
\label{eq:att_sens}
    \psi_{ATT} \in \left[\frac{\E[AY^1] - \delta\E[\pi(X)\mu_0(X)]}{\E[A]}, ~\frac{\E[AY^1] - \frac{1}{\delta}\E[\pi(X)\mu_0(X)]}{\E[A]} \right].
\end{equation}
\end{proposition}

By relaxing the NUC assumption we can no longer point identify the causal parameter of interest $\psi_{ATT}$. Rather, we can identify \textit{causal bounds} for $\psi_{ATT}$. We proceed as before and use semiparametric theory to propose estimators for these bounds using one-step estimators. To our knowledge, this is the first derivation of a risk-ratio-based sensitivity analysis for the ATT when the treatment and outcome are not always observed.


\begin{lemma}
\label{lema:dr_bounds}
    Assume (A1)-(A2), (R1)-(R2) and (\ref{eq:sensitivity_ratio}) hold. Then 
    \begin{equation}
    \label{eq:dr_bounds}
        \widehat{L}_{ATT} = \frac{\widehat{\psi}_{ay^1} - \delta\widehat{\psi}_{ay^0}}{\widehat{\psi}_{a}}, ~~~~~ \widehat{U}_{ATT} = \frac{\widehat{\psi}_{ay^1} - \frac{1}{\delta}\widehat{\psi}_{ay^0}}{\widehat{\psi}_{a}}
    \end{equation}
    are one-step estimators for the causal lower and upper bounds in \eqref{eq:att_sens}, where $\widehat{\psi}_{ay^1}, \widehat{\psi}_{ay^0}$ and $\widehat{\psi}_{a}$ are given by \eqref{eq:onestep_comp}.
\end{lemma}

Finally, to calibrate the unmeasured confounding to the measured controls observed in our data we estimate the ratio (\ref{eq:sensitivity_ratio}) using only a subset of the covariates $V \subset X$. That is, we assume that the numerator is not identified under $V$, but it is identified under $X$, and use ratio \eqref{eq:sensitivity_ratio} to measure the bias introduced by the confounders in $X$ but not in $V$:
\begin{equation}
\label{eq:ident_ratio}
    \frac{\E[Y^0 | V, A=1, R=1]}{\E[Y^0 | V, A=0, R=1]} = \frac{\E[\E[Y|X,A=0,R=1] | V, A=1, R=1]}{\E[Y | V, A=0, R=1]} \leq \delta.
\end{equation}

\section*{Estimation}

In the data analysis, we flexibly estimate the nuisance functions $\pi(X) = \E[A \mid X,R=1]$, $\omega(X) = \PP(R=1 \mid X)$, $\mu_1(X) = \E[AY \mid X,R=1]$ and $\mu_0(X) = \E[Y \mid X,R=1,A=0]$ using the SuperLearner methodology (\cite{superlearner}) with Random Forest and Logistic regression as base models, and 10-fold sample splitting. We then follow the logic of the \texttt{npcausal} library\footnote{\url{https://github.com/ehkennedy/npcausal}.} implemented in R to estimate the corresponding one-step estimators of interest \eqref{eq:onestep_comp}, and finally \eqref{eq:onestep_att}.

Concretely, we divide our dataset into $K=10$ folds of (approximately) equal size. Each of these folds is set aside while the other 9 folds are used to estimate the nuisance functions $\pi, \omega, \mu_1, \mu_0$. These trained models are then used to produce out-of-sample predictions on the hold-out fold for $\widehat{\varphi}_{ay^1,k}, \widehat{\varphi}_{a,k}, \widehat{\varphi}_{ay^0,k}$. With these predictions, the $\widehat{\psi}_{ATT, k}$ estimator is constructed for each fold $k$, following \eqref{eq:onestep_att}. The average of $\widehat{\psi}_{ATT, k}$ for the 10 folds is the final estimate for $\widehat{\psi}_{ATT}$.

\section{Results: Effect of victimization on future offending among adolescents}
\label{sec:results}

In this section, we present and analyze the results found by applying the developed methodology detailed in Section \ref{sec:methods} to the Add Health dataset described in Section \ref{sec:data}. We first present the ATT of victimization on future offending among adolescents. We then investigate the existence of heterogeneous effects according to the age of the victimization. Next, we perform a sensitivity analysis on our NUC assumption under different confounding levels. Finally, we disaggregate our results between the effect of victimization on violent vs. non-violent offending.

\section*{Average Treatment effect on the Treated}

We estimate how the offending rate among victimized individuals would have changed if none of them had been victimized:
\begin{equation*}
    \widehat{\psi}_{ATT} = \widehat{\E}[Y^1 - Y^0 | A=1].
\end{equation*}
We estimate $\widehat{\psi}_{ATT}$ using the doubly robust estimator derived in Theorem \ref{thm:onestep_att} in Section \ref{sec:methods}.

We find a positive effect of victimization on future offending among adolescents. Our results show that the offending rate among victimized individuals would have been 3.86 percentage points lower ($95\%$ CI: $[0.28, 7.45]$) if none of them had been victimized. In our sample, our estimations imply that the offending rate would have decreased from $11.51\%$ to $7.64\%$ among victimized individuals. This represents a reduction of $33.5\%$ in the future offending rate among victimized adolescents ($95\%$ CI: $[2.4\%, 64.7\%]$). 

The Appendix discusses the results using alternative causal parameters of interest, such as the Average Treatment Effect (ATE) and the Overall Treatment Removal effect (OTR). We believe that the ATE is less informative than the ATT in this context since the latter allows us to study what would happen if those who are currently victimized were not victimized, shedding light on the benefits of public policies that seek to reduce adolescent victimization, rather than comparing two hypothetical scenarios in which all vs. none are victimized. Additionally, identifying $\E[Y^1]$ requires the additional positivity assumption $\PP(A=1|X) \geq \epsilon > 0$ w.p. 1, yet in our data set our estimates show that there are levels of the covariates for which the probability of being victimized is negligible. The results using the OTR are consistent with the ones using the ATT.
 
\section*{Heterogeneous ATT by age}

\begin{figure}[h]
    \centering
    \includegraphics[width=.6\textwidth]{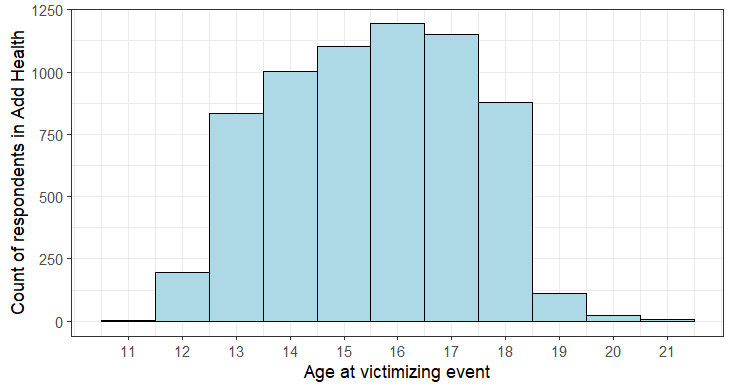}
    \caption{Distribution of age at Wave 1 in the Add Health dataset.}
    \label{fig:ages}
\end{figure}

In addition to the overall ATT of victimization on future offending among adolescents, we are also interested in exploring if the estimated ATT varies as a function of the age at which the victimizing event is experienced. To this end, we estimate heterogeneous ATT effects by age using the observed age in Add Health collected at Wave 1. Recall treatment is measured one year later at Wave 2 by asking respondents whether during the previous 12 months they experienced a victimizing event. Figure \ref{fig:ages} details the number of observations in each age group for the 6,504 respondents in Add Health dataset. 



\begin{figure}
    \centering
    \includegraphics[width=.66\textwidth]{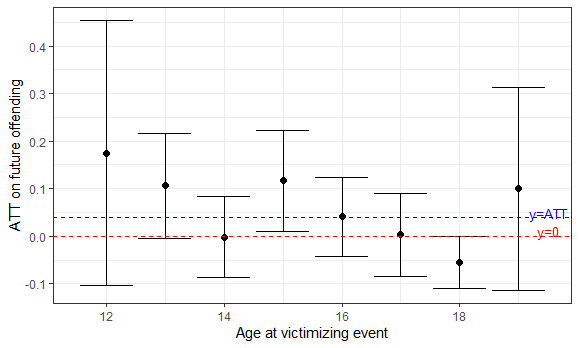}
    \caption{Heterogeneous ATT of victimization on future offending by age at the victimizing event using doubly robust estimators.}
    \label{fig:att_age}
\end{figure}

We group respondents with $\leq 12$ years old and $\geq 19$ years old to avoid uninformative confidence intervals due to the small number of observations in these categories. For this analysis, we ignore three observations with missing age information. The ATT estimates and their standard deviations are estimated by taking the sample average and variance of the one-step estimator $\psi_{ATT,g}$ for each age group, according to equation (\ref{eq:onestep_att}) and the variance results in Corollary \ref{cor:normal_att}. Figure \ref{fig:att_age} plots the point estimate for the ATT and its standard deviation of victimization on future offending for each age group. 

Our results suggest that there are heterogeneous effects depending on the age at the victimizing event. Specifically, Figure \ref{fig:att_age} shows a decreasing ATT by age group. However, only for 15-year-olds, it can be ruled out that the effect of victimization on offending is zero. We formally test for heterogeneous effects by testing the null hypothesis that all the effects are simultaneously equal using the doubly robust test proposed in \cite{heterogeneous_ehk}. A detailed explanation of the test is given in Section E of the Appendix. We obtain a $p$-value of $0.039$ providing statistical evidence to reject the null hypothesis of homogeneous ATT by age group. 

Hence, we can conclude that the effect of victimization on future offending varies depending on the age at the victimizing event. This is one of the main motivations of our work as we hypothesize that the cycle of violence theory, widely studied for abused children, should not be directly extended to older populations. In more detail, our results suggest that the effect of victimization on future delinquency among those victimized decreases as the victimizing event is experienced at an older age, even becoming negative (but not statistically significant) by age 18. This behavior is explained, for example, by developmental psychopathology or life course theories (\cite{unsafe_mersky}).

\section*{Sensitivity analysis to NUC assumption}

In the following, we stress test our assumption of No Unmeasured Confounders (A3) to assess the robustness of our results to missing relevant confounders. This is crucial if there are omitted variables correlated at the same time with both the likelihood of experiencing victimization events and that of future offending. In the presence of such unmeasured confounders, our results may be biased. Therefore, we are interested in understanding how our results would change if this were the case.

Our relaxed assumption \ref{eq:sensitivity_ratio} captures the intuition that there may be unmeasured confounders, where the offending rate of victimized individuals can be different from those who were not victimized, even if victimized respondents had not experienced  victimization. For instance, adolescents with certain unobserved characteristics (e.g., cultural norms) may be more likely to be exposed to victimization events and, at the same time, may be more likely to engage in criminal activities, even when they are not victimized. 

\begin{figure}
    \centering
    \includegraphics[width=.65\textwidth]{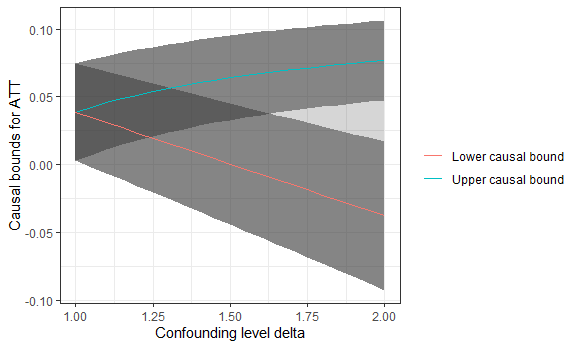}
    \caption{Sensitivity analysis for the No Unmeasured Confounders assumption using the causal bounds given by Proposition \ref{prop:att_sens} and doubly robust estimators (Lemma \ref{lema:dr_bounds}).}
    \label{fig:sens_att}
\end{figure}

Figure \ref{fig:sens_att} presents the results of our sensitivity analysis. The upper and lower causal bounds derived in Proposition \ref{prop:att_sens} are plotted in blue and red, respectively. These are estimated using the doubly robust estimators given by Lemma \ref{lema:dr_bounds}. These estimators have their own confidence intervals constructed analogously to equation (\ref{eq:ci_dr}) for $\widehat{\psi}_{ATT}$ and are presented in dark gray for each bound. Finally, the light gray represents the possible values of $\psi_{ATT}$. The parameter $\delta$ is varied to study the extent to which our results change if the NUC assumption holds ($\delta = 1$) and under unmeasured confounders up to twice as important as those included in our estimation ($\delta = 2$). 

The sensitivity analysis shows that our results hold for $\delta < 1.05$. That is, our results are robust if there are no unmeasured confounders, correlated with both victimization and future offending, such that the offending rate of victimized individuals had they not been victimized is 5\% greater than the offending rate of those not victimized.

\begin{figure}
    \centering
    \includegraphics[width=.6\textwidth]{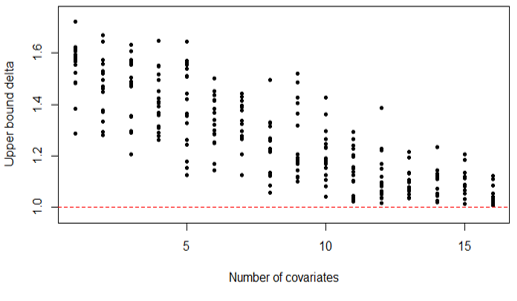}
    \caption{Estimation of confounding bias using different subsets of the available covariates \ref{eq:ident_ratio}}
    \label{fig:sens_delta}
\end{figure}

To have a sense of plausible values for $\delta$, we simulate the situation in which we do not observe all the available covariates, but a subset of them, and estimate the bias introduced by such omitted variables. In more detail, we use random subsets of variables $V$ and estimate the bias introduced by the confounding variables in $X$ but not in $V$. This is a way of calibrating the sensitivity anlysis to observed measured confounding. Figure \ref{fig:sens_delta} plots the estimated ratio (\ref{eq:ident_ratio}) using this approach and doubly robust estimators as an estimate of the upper bound $\delta$. As expected, the estimated upper bound $\delta$ decreases as we include more variables in the estimation. The estimated $\delta \in [1.006, ~ 1.721]$, together with Figure \ref{fig:sens_att}, suggests that if the unmeasured confounders bias our results as much as some (small) subsets of the available covariates, such that $\delta < 1.05$, then our results of a positive causal relationship between victimization and future offending for victimized adolescents are robust.

\section*{ATT of victimization on violent vs. non-violent offending} 

Finally, we break down the found effect of victimization in adolescents between the effect on violent and non-violent offenses. To this end, we replicate our analysis varying the definition of the outcome of interest $Y$ to include exclusively either violent or non-violent crimes.

In detail, we define non-violent offending as whether in the 12 months prior to Wave 3 the respondent \textit{stole something worth more than \$50} or \textit{broke into a house or building to steal something (burglary)}. Using this definition, we find a larger effect showing that the offending rate of non-violent crimes would have been $4.33$ percentage points lower (95\% CI: $[1.72, 6.96]$) if none of the victimized adolescents had been victimized. 

Furthermore, we find statistical evidence of a heterogeneous ATT of victimization on engagement in non-violent crimes in the future according to the age at which the victimizing event is experienced. Using the doubly robust test given by equation (\ref{eq:test_age}), we obtain a $p$-value of $0.0058$ and a decreasing effect by age group. Finally, the sensitivity analysis shows these results are again robust to modest unmeasured confounding, up to $\delta \leq 1.1$.

On the other hand, we define violent offending as whether the subject \textit{pointed a knife or gun at someone, shot or stabbed someone}, or \textit{used or threatened to use a weapon to get something from someone (robbery)} in the 12 months prior to Wave 3. However, we find that the effect of victimization on violent offending is not statistically significant ($\widehat{\psi}_{ATT} = 0.99$, 95\% CI: $[-1.99, 3.97]$). 

We conclude that the effect found is mainly driven by non-violent crimes, such as minor thefts and burglary, rather than by violent crimes.

\section{Discussion}
\label{sec:discussion}

We found evidence for a positive causal effect of victimization on future offending among adolescents using the Add Health dataset. We found that, among victimized respondents, if none of them had been victimized and all had been observed, the delinquency rate would have been $3.86$ percentage points lower ($95\%$ CI: $[0.28, 7.45]$). Considering that the estimated offending rate among victimized individuals is $11.51\%$, this represents a reduction of $33.5\%$ in the future offending rate ($95\%$ CI: $[2.4\%, 64.7\%]$). One important contribution of our work is the derivation of a doubly robust influence-function-based estimator for the ATT when the treatment and outcome are partially missing.

We further investigated whether this effect varies according to the age at which the victimizing event is experienced. Using a doubly robust test, we were able to reject the null hypothesis of homogeneous effects, that is, the effect of victimization on future offending does vary depending on the age at which the victimization occurs. This confirms one of our main hypotheses regarding the validity of the cycle of violence theory in other age groups. In particular, our results show that the effect of victimization on future offending decreases with age, as predicted by developmental psychopathology and life course theories.

Moreover, we derived a novel risk-ratio-based sensitivity analysis in this setting to assess the robustness of our results to unmeasured confounders. We found that our results are robust to modest unmeasured confounding such that the ratio between the potential outcome of victimized individuals, had they not been victimized, and the offending rate of those not victimized is less than $1.05$.

For instance, a relevant confounding factor in our setting is previous victimization or offending events, as these are positively correlated with both facing new victimization experiences and with violence later in life. However, given that Add Health began collecting data when respondents were adolescents, we cannot be sure that they did not suffer victimization, for instance, during their childhood.  We try to test how this confounder affects our results by re-estimating the ATT of victimization at Wave 2 on offending at Wave 3, restricting our sample to those with no prior record of victimization or delinquency in Add Health. 

In detail, we estimate the ATT for those who (i) were not victimized at Wave 1, (ii) did not offend at Wave 1, (iii) did not offend at Wave 2, and (iv) were not sexually abused before sixth grade by their parents or other adult caregivers. Note that our main results include victimization and offending events in Wave 1 as control variables. Among this subgroup, we found a negligible ATT effect of victimization on future offending of $-0.82$ percentage points non-statistically significant ($95\%$ CI: $[-4.67, 3.03]$). However, this estimation is likely still confounded by previous victimization events not captured by Add Health survey, which can bias the results in an unknown way. 

Overall, 
there is statistical evidence of an unfavorable causal relationship between victimization and future offending among adolescents. This effect is mainly driven by non-violent offending and decreases by the age at the victimizing event. Our results are robust to modest confounding, but sensitive to the specific methodological design. Future work should be done to address important confounders such as previous victimization events and the interdependence of victimization and offending throughout life.


\bibliographystyle{imsart-nameyear}
\bibliography{bibliography.bib}

\newpage
\section*{Appendix}

\subsection*{A} \textit{Descriptive statistics of control variables.}

\begin{table}[H]
\caption{Descriptive statistics of categorical control variables}
\begin{tabular}{lcc}
\toprule
\multicolumn{1}{c}{Variable}           & Description                              & \% Missing \\
\midrule
Sex                                    & 51.6\% Female, 48.4\% Male               & 0.02       \\
Race                                   & 22.5\% Black, 60.7\% White, 16.7\% Other & 0.10       \\
Born in the US                         & 93.8\% Yes, 6.1\% No                     & 0.08       \\
Currently working                      & 56.7\% Yes, 42.8\% No                    & 0.40       \\
Presently in school                    & 98.03\% Yes, 1.92\% No                   & 0.05       \\
Marital status (had been married ever) & 0.54\% Yes, 99.43\% No                   & 0.03       \\
Repeated a grade                       & 21.48\% Yes, 78.31\% No                  & 0.21       \\
Should get medical care but did not    & 19.21\% Yes, 80.60\% No                  & 0.18       \\
Considered suicide                     & 12.62\% Yes, 86.32\% No                  & 1.04       \\
Got drunk more than once a month       & 9.86\% Yes, 89.91\% No                   & 0.23       \\
Illicit drug use                       & 29.46\% Yes, 69.45\% No                  & 1.09       \\
People care in neighborhood            & 72.16\% Yes, 25.62\% No                  & 2.23       \\
Feel safe in neighborhood              & 89.07\% Yes, 10.38 \% No                 & 0.55       \\
Parents receive welfare                & 7.90\% Yes, 78.40\% No                   & 13.70      \\
Parents have enough money              & 69.13\% Yes, 15.51\% No                  & 15.36      \\
Victimization at Wave 1                & 10.06\% Yes, 79.27\% No                  & 0.77       \\
Offending at Wave 1                    & 12.67\% Yes, 86.47\% No                  & 0.86      \\
\bottomrule
\end{tabular}
\end{table}

\begin{table}[H]
\caption{Descriptive statistics of continuous control variables}
\begin{tabular}{lcccc}
\toprule
\multicolumn{1}{c}{Variable}                                  & Range        & Mean  & Std  & \% Missing \\
\midrule   
Age                                       & [11, 21] & 15.53 & 1.78 & 0.05       \\
Self-rated health                         & [1, 5]   & 3.90  & 0.90 & 0.12       \\
Parental supervision                      & [0, 1]   & 0.74  & 0.22 & 2.57       \\
Self-control                              & [0, 4]   & 1.04  & 0.73 & 2.12       \\
Perceived importance of post-HS education & [1, 5]   & 4.44  & 1.02 & 0.81       \\
Expectation of post-HS education          & [1, 5]   & 4.19  & 1.20 & 0.00 \\      
\bottomrule
\end{tabular}
\end{table}

\subsection*{B} \textit{Average Treatment Effect}. The Average Treatment Effect $\E[Y^1 - Y^0]$ is a causal parameter commonly employed to estimate the effect of a treatment (e.g. victimization) on an outcome (e.g. offending). The ATE asks \textit{what would have happened if all vs. none had been victimized}. However, in our setting, the potential outcome had all respondents been victimized is not of interest from a public policy perspective. This is an extreme counterfactual that considers the situation in which all respondents are victimized, which is an undesirable and unrealistic scenario.

\begin{figure}
    \centering
    \includegraphics[width=.65\textwidth]{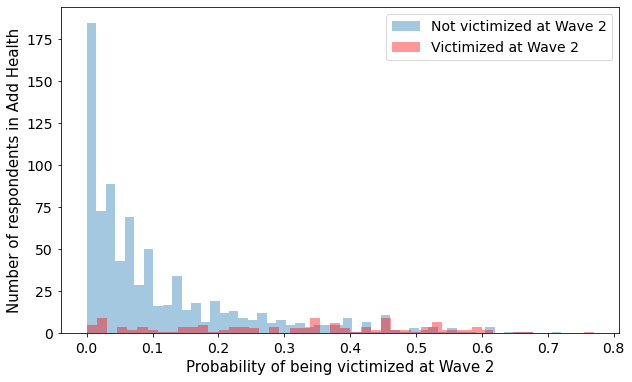}
    \caption{Propensity score of victimization at Wave 2 given covariates on test set using Random Forest.}
    \label{fig:ps_victim}
\end{figure}

Furthermore, to identify $\E[Y^1]$ and estimate it property with data we would need the additional causal assumption that $\PP(A=1 | X) > 0$. That is, that everyone in our population has some positive probability of being victimized. Nonetheless, a considerable portion of Add Health respondents have negligible probabilities of victimization given their demographic and socio-economic attributes considered, as is shown in Figure \ref{fig:ps_victim}. This is an expected phenomenon, as the attributes of some respondents ``protect'' them from experiencing victimization events.   

\subsection*{C} \textit{Overall Treatment Removal effect.} The Overall Treatment Removal effect (OTR) asks how the offending rate would have changed in our sample if none of the individuals had been victimized. Mathematically, we are interested in estimating the following causal parameter given our observed data
\begin{equation}
    \psi_{OTR} = \E\left[Y\right] - \E\left[Y^{0}\right].
\end{equation}

Under equivalent assumptions used for the ATT, we have the following one-step estimator for $\psi_{OTR}$ (\cite{semiparametric})
\begin{equation}
\begin{split}
     \widehat{\psi}_{OTR} = \widehat{\E}[Y] - \widehat{\E}[Y^0] = & \PP_n\left[\frac{R}{\widehat{\omega}(X)}\left(Y - \widehat{\E}(Y|R=1, X)\right) + \widehat{\E}(Y|R=1, X) \right] \\
     & - \PP_n\left[\frac{R(1-A)\widehat{\pi}(X)}{\widehat{\omega}(X)(1-\widehat{\pi}(X))}(Y-\widehat{\mu}_0(X)) + \widehat{\mu}_0(X) \right],
\end{split}
\end{equation}
where $\omega(X) = \PP(R=1|X)$, $\pi(X) = \E(A|X)$, $\mu_0(X) = \E[Y|X,R=1,A=0]$ and $\PP_n(Z)$ is the sample average of $Z$, $\frac{1}{n}\sum_{i=1}^{n} Z_i$.

We found a positive overall treatment removal effect suggesting that being victimized as an adolescent has a causal effect on engaging in criminal activities in the future. Specifically, if none of the respondents had been victimized, the delinquency rate would have been 0.67 percentage points lower ($95\%$ CI: $[0.19, 1.16]$), which represents a reduction of about 10\% in the offending rate.

\subsection*{D} \textit{Proofs}

\begin{proof}[Proof of Proposition \ref{prop:att_identification}]

We rewrite the ATT as
\begin{equation}
    \psi_{ATT} = \E[Y^1-Y^0|A=1] = \frac{\E[A(Y^1-Y^0)]}{\PP(A=1)} = \frac{\E[AY^1]-\E[AY^0]}{\E[A]},
\end{equation}
and identify each term independently.

First, the Victimization rate $\E[A]$ in the denominator needs to be identified due to attrition not completely at random. Under our causal assumptions, we have 
\begin{equation}
    \E[A]= \E[\E[A|X]] = \E[\E[A|X, R=1]],
\label{eq:ident_a}
\end{equation}
using iterated expectations, and then (R1)-(R2) assumptions. 

Next, the potential outcome for victimized individuals had they been victimized $\E[AY^1]$, the first term in the numerator, needs to be identified due to dropout not completely at random. Following similarly we have
\begin{equation}
    \E[AY^1] = \E[AY] = \E[\E[AY|X]] = \E[\E[AY|X, R=1]].
\label{eq:ident_ay1}
\end{equation}
The first equality uses Consistency (A1), the second uses iterated expectations, and the last Missing at Random (R1)-(R2) assumptions. 

The remaining term in the numerator, the potential outcome for victimized individuals had they not been victimized $\E[AY^0]$, needs to be identified due to dropout not being completely at random and because we do not observe this potential outcome for victimized individuals as they are indeed victimized. Using our causal assumptions we can identify $\E[AY^0]$ with the observed data by
\begin{equation}
\begin{split}
    \E[AY^0] & = \E[\E[AY^0|X]] = \E[\E[A|X]\E[Y^0|X]] \\
    & = \E[\E[A|X,R=1]\E[Y^0|X,R=1,A=0]] \\
    & = \E[\E[A|X,R=1]\E[Y|X,R=1,A=0]],
    \label{eq:ident_ay0}
\end{split}
\end{equation}
where we used iterated expectations, then independence given the No Unmeasured Confounders assumption (A3), then (A2)-(A3) + (R1)-(R2), and finally Consistency (A1).

Putting the pieces together, we obtain the desired identifying result
\begin{equation}
    \psi_{ATT} = \frac{\E[\E[AY|X, R=1]] - \E[\E[A|X,R=1]\E[Y|X,R=1,A=0]]}{\E[\E[A|X, R=1]]}.
\end{equation}

\end{proof}

\begin{proof}[Proof of Lemma \ref{lema:att_reminders}]

Using standard techniques (\cite{ehk_semiparametric}) we derive the (centered) efficient influence function for $\psi_{ay^0} = \E[\E[A|X,R=1]\E[Y|X,R=1,A=0]]$:
$$\varphi_{ay^0} = \frac{R}{\widehat{\omega}(X)}(A-\widehat{\pi}(X))\widehat{\mu}_0(X) + \frac{R}{\widehat{\omega}(X)}\frac{(1-A)\widehat{\pi}(X)}{1-\widehat{\pi}(X)}(Y-\widehat{\mu}_0(X)) + \widehat{\pi}(X)\widehat{\mu}_0(X) - \psi_{ay^0}.$$

    Then, we can write $\widehat{\psi}_{ay^0} - \psi_{ay^0}$ as the sum of three terms given that $\widehat{\psi}_{ay^0}$ is the sample average of an estimated function $\widehat{\psi}_{ay^0} = \PP_n(\widehat{\varphi}_{ay^0}) + \psi_{ay^0}(\widehat{\PP})$:
    $$\widehat{\psi}_{ay^0} - \psi_{ay^0} =  (\PP_n - \PP)\left\{\varphi_{ay^0}\right\} + (\PP_n - \PP)\left(\widehat{\varphi}_{ay^0} - \varphi_{ay^0}\right) + \PP\left(\widehat{\varphi}_{ay^0} - \varphi_{ay^0}\right).$$

    The second term was shown by \cite{ehk2020sharp} to be 
    $$(\PP_n - \PP)\left(\widehat{\varphi}_{ay^0} - \varphi_{ay^0}\right) = O_{\PP}\left(\frac{\norm{\widehat{\varphi}_{ay^0} - \varphi_{ay^0}}}{\sqrt{n}}\right) = o_{\PP}(1/\sqrt{n}),$$
    using sample splitting and given that $\norm{\widehat{\varphi}_{ay^0} - \varphi_{ay^0}} = o_{\PP}(1)$ by assumption. 
    
    Setting $R_{2,ay^0} = \PP(\widehat{\varphi}_{ay^0} - \varphi_{ay^0})$ we obtain the desired result. Therefore, we only need to bound the third term $R_{2,ay^0} = \PP(\widehat{\varphi}_{ay^0} - \varphi_{ay^0})$. In detail, dropping the $X$ argument to ease notation, we have
    \begin{align*}
        \PP(\widehat{\varphi}_{ay^0} - \varphi_{ay^0}) & = \E\left[\frac{R}{\widehat{\omega}}(A-\widehat{\pi})\widehat{\mu}_0 + \frac{R}{\widehat{\omega}}\frac{(1-A)\widehat{\pi}}{1-\widehat{\pi}}(Y-\widehat{\mu}_0) + \widehat{\pi}\widehat{\mu}_0 - \pi \mu_0\right] \\
        & = \E\left[\frac{\omega}{\widehat{\omega}}(\pi-\widehat{\pi})\widehat{\mu}_0 + \frac{\omega}{\widehat{\omega}}\frac{(1-\pi)\widehat{\pi}}{1-\widehat{\pi}}(\mu_0-\widehat{\mu}_0) + \widehat{\pi}\widehat{\mu}_0 - \pi \mu_0\right] \\
        & = \E\left[\frac{\omega}{\widehat{\omega}}\left((\pi-\widehat{\pi})(\widehat{\mu}_0 - \mu_0) + (\pi-\widehat{\pi})\mu_0 + \frac{(1-\pi)\widehat{\pi}}{1-\widehat{\pi}}(\mu_0-\widehat{\mu}_0)\right) + \widehat{\pi}\widehat{\mu}_0 - \pi \mu_0\right] \\
        & = \E\left[\frac{\omega}{\widehat{\omega}}\left((\pi-\widehat{\pi})\mu_0 + (\mu_0-\widehat{\mu}_0)\left\{\frac{(1-\pi)\widehat{\pi}}{1-\widehat{\pi}}-(\pi-\widehat{\pi})\right\}\right) + \widehat{\pi}\widehat{\mu}_0 - \pi \mu_0\right] \\
        & = \E\left[\frac{\omega}{\widehat{\omega}}\left((\pi-\widehat{\pi})\mu_0 + (\mu_0-\widehat{\mu}_0)\left\{\frac{\widehat{\pi}-\pi}{1-\widehat{\pi}} + \widehat{\pi}\right\}\right) + \widehat{\pi}\widehat{\mu}_0 - \pi \mu_0\right] \\
        & = \E\left[\frac{\omega}{\widehat{\omega}}\left((\mu_0-\widehat{\mu}_0)\frac{\widehat{\pi}-\pi}{1-\widehat{\pi}} + \pi\mu_0 - \widehat{\pi}\widehat{\mu}_0\right) + \widehat{\pi}\widehat{\mu}_0 - \pi \mu_0\right] \\
        & = \E\left[\frac{\omega}{\widehat{\omega}}(\mu_0-\widehat{\mu}_0)\frac{\widehat{\pi}-\pi}{1-\widehat{\pi}} + (\pi\mu_0 - \widehat{\pi}\widehat{\mu}_0) \frac{\omega- \widehat{\omega}}{\widehat{\omega}}\right] \\
        & \lesssim  \norm{\widehat{\pi} - \pi}\norm{\widehat{\mu}_0 - \mu_0} + \norm{\widehat{\omega} - \omega}(\norm{\widehat{\pi} - \pi} + \norm{\widehat{\mu}_0 - \mu_0}),
    \end{align*}
    as claimed. The first equality uses iterated expectations and the definition of the nuisances functions $\omega(X) = \PP[R=1|X]$, $\pi(X) = \E[A|X, R=1]$, and $\mu_0(X) = \E[Y|X, R=1, A=0]$. Moreover, the final inequality uses Cauchy-Schwarz and the assumption that $\widehat{\pi}$ and $\widehat{\omega}$ are bounded away from 0 and 1 with probability one.
    
\end{proof}

\begin{proof}[Proof of Theorem \ref{thm:onestep_att}]

\begin{align*}
    \widehat{\psi}_{ATT} - \psi_{ATT} & = \frac{\widehat{\psi}_{ay^1} - \widehat{\psi}_{ay^0}}{\widehat{\psi}_{a}} - \frac{\psi_{ay^1} - \psi_{ay^0}}{\psi_{a}} \\
    & = \frac{\widehat{\psi}_{ay^1} - \psi_{ay^1}}{\widehat{\psi}_{a}} - \frac{\widehat{\psi}_{ay^0} - \psi_{ay^0}}{\widehat{\psi}_{a}} + (\psi_{ay^1} - \psi_{ay^0})\left(\frac{1}{\widehat{\psi}_{a}} - \frac{1}{\psi_{a}}\right) \\
    & = \frac{\widehat{\psi}_{ay^1} - \psi_{ay^1}}{\widehat{\psi}_{a}} - \frac{\widehat{\psi}_{ay^0} - \psi_{ay^0}}{\widehat{\psi}_{a}} + \left(\frac{\psi_{ay^1} - \psi_{ay^0}}{\psi_{a}}\right)\left(\frac{\psi_{a} - \widehat{\psi}_{a}}{\widehat{\psi}_{a}}\right) \\
    & = \frac{1}{\widehat{\psi}_{a}}\left((\widehat{\psi}_{ay^1} - \psi_{ay^1}) - (\widehat{\psi}_{ay^0} - \psi_{ay^0}) \right) - \psi_{ATT}\left(\frac{\psi_{a} - \widehat{\psi}_{a}}{\widehat{\psi}_{a}}\right) \\
    & = \frac{1}{\psi_{a}}\left((\widehat{\psi}_{ay^1} - \psi_{ay^1}) - (\widehat{\psi}_{ay^0} - \psi_{ay^0}) - \psi(\widehat{\psi}_{a} - \psi_{a}) \right) \\
    & ~~~~~~ + \left(\frac{1}{\widehat{\psi}_{a}} - \frac{1}{\psi_{a}}\right) \left((\widehat{\psi}_{ay^1} - \psi_{ay^1}) - (\widehat{\psi}_{ay^0} - \psi_{ay^0}) - \psi(\widehat{\psi}_{a} - \psi_{a}) \right).
\end{align*}

Next, we can use Lemma \ref{lema:att_reminders} to analyze the term
\begin{align*}
    (\widehat{\psi}_{ay^1} & - \psi_{ay^1}) - (\widehat{\psi}_{ay^0} - \psi_{ay^0}) - \psi(\widehat{\psi}_{a} - \psi_{a}) = \\
    & = (\PP_n - \PP)\left\{\varphi_{ay^1} \right\} + o_{\PP}(1/\sqrt{n}) + R_{2,ay^1} - \left((\PP_n - \PP)\left\{\varphi_{ay^0} \right\} + o_{\PP}(1/\sqrt{n}) + R_{2,ay^0} \right) \\
    & ~~~~~ -\psi\left((\PP_n - \PP)\left\{\varphi_{a} \right\} + o_{\PP}(1/\sqrt{n}) + R_{2,a}\right) \\
    & = (\PP_n - \PP)\left\{\varphi_{ay^1} - \varphi_{ay^0} - \psi \varphi_{a} \right\} + o_{\PP}(1/\sqrt{n}) + R_{2,ay^1} - R_{2,ay^0} - \psi R_{2,a}.
\end{align*}

The first term in this expression is a centered sample average. Thus, by the Central Limit Theorem, $(\PP_n - \PP)\left\{\varphi_{ay^1} - \varphi_{ay^0} - \psi \varphi_{a} \right\} = O_{\PP}(1/\sqrt{n}).$ In addition, given our third and fourth assumptions,
$\frac{1}{\widehat{\psi}_{a}} - \frac{1}{\psi_{a}} = o_{\PP}(1)$. Therefore,

\begin{align*}
    \widehat{\psi}_{ATT} - \psi_{ATT} & = \frac{1}{\psi_{a}}\left((\PP_n - \PP)\left\{\varphi_{ay^1} - \varphi_{ay^0} - \psi \varphi_{a} \right\} + o_{\PP}(1/\sqrt{n}) + R_{2,ay^1} - R_{2,ay^0} - \psi R_{2,a}\right) \\
    & ~~~~~~ + o_{\PP}(1)\left(O_{\PP}(1/\sqrt{n}) + o_{\PP}(1/\sqrt{n}) + R_{2,ay^1} - R_{2,ay^0} - \psi R_{2,a}\right) \\
    & = (\PP_n - \PP)\left\{\varphi_{ATT} \right\} + o_{\PP}(1/\sqrt{n}) + (R_{2,ay^1} - R_{2,ay^0} - \psi R_{2,a})\left(1 + o_{\PP}(1)\right),
\end{align*}
given the definition of $\varphi_{ATT}$ and that $o_{\PP}(1/\sqrt{n}) + o_{\PP}(1)(O_{\PP}(1/\sqrt{n}) + o_{\PP}(1/\sqrt{n})) = o_{\PP}(1/\sqrt{n}) + o_{\PP}(1)O_{\PP}(1/\sqrt{n}) = o_{\PP}(1/\sqrt{n}) + o_{\PP}(1/\sqrt{n}) = o_{\PP}(1/\sqrt{n}).$

To finish our proof we need to bound the remainder term. Given Lemmas \ref{lemma:onestep_ay1}, \ref{lemma:onestep_a} and \ref{lemma:onestep_ay0}, we have
\begin{align*}
    R_{2,ay^1} - R_{2,ay^0} - \psi R_{2,a}  & \lesssim \norm{\widehat{\omega} - \omega}\norm{\widehat{\mu}_1 - \mu_1} + \norm{\widehat{\omega} - \omega}(\norm{\widehat{\pi} - \pi} + \norm{\widehat{\mu}_0 - \mu_0}) \\
    & ~~~~~ + \norm{\widehat{\pi} - \pi}\norm{\widehat{\mu}_0 - \mu_0} + \norm{\widehat{\omega} - \omega}\norm{\widehat{\pi} - \pi} \\
    & = \norm{\widehat{\omega} - \omega}(\norm{\widehat{\pi} - \pi} + \norm{\widehat{\mu}_0 - \mu_0} + \norm{\widehat{\mu}_1 - \mu_1}) + \norm{\widehat{\pi} - \pi}\norm{\widehat{\mu}_0 - \mu_0},
\end{align*}

\end{proof}

\begin{proof}[Proof of Proposition \ref{prop:att_sens}]
Let us write the causal parameter of interest $\psi_{ATT}$ as
\begin{equation}
\label{eq:att_exp}
    \psi_{ATT} = \E[Y^1-Y^0|A=1] = \frac{\E[AY^1]-\E[AY^0]}{\E[A]}.
\end{equation}

Next, we use iterated expectations together with (R1) and (R2) to note that
$$\E[AY^0] = \E[\E[AY^0 | X]] = \E[\E[AY^0 | X, R=1]],$$
and manipulate the term inside the expectation $\E[AY^0| X, R=1]$. Given our assumptions, we have
\begin{equation}
\begin{split}
\label{eq:ay0_sensitivity}
    \E[AY^0 | X, R=1] 
    & = \E[\E[AY^0 | X, A=1, R=1] | X, R=1] \\
    & = \E[A\E[Y^0 | X, A=1, R=1] | X, R=1] \\
    & = \E[A | X, R=1] \E[Y^0 | X, A=1, R=1] \\
    & = \E[A | X, R=1] \E[Y^0 | X, A=0, R=1] \frac{\E[Y^0 | X, A=1, R=1]}{\E[Y^0 | X, A=0, R=1]}  \\
    & = \pi(X)\mu_0(X) \frac{\E[Y^0 | X, A=1, R=1]}{\E[Y^0 | X, A=0, R=1]}.
\end{split}
\end{equation}

Therefore, 
\begin{equation}
    \frac{1}{\delta} \pi(X)\mu_0(X) \leq E[AY^0 | X, R=1] \leq \delta \pi(X)\mu_0(X),
\end{equation}
and
\begin{equation}
    \E[AY^0] = \E[\E[AY^0 | X, R=1]] \in \left[\frac{1}{\delta} \E[\pi(X)\mu_0(X)], ~ \delta \E[\pi(X)\mu_0(X)] \right]. 
\end{equation}

Plugging this in equation (\ref{eq:att_exp}) gives the desired result.
\end{proof}

\begin{proof}[Proof of Lemma \ref{lema:dr_bounds}]
    The proof is analogous to that of Theorem \ref{thm:onestep_att} using Lemma \ref{lema:att_reminders}.
\end{proof}

\subsection*{E} \textit{Heterogeneous treatment effects}

We formally test for heterogeneous effects by testing the null hypothesis that all the effects are simultaneously equal:
\begin{equation}
\label{eq:h0_age}
    H_0: \widehat{\psi}_{j} = \widehat{\psi}_k, ~~\forall~j,k \in \{age\}.
\end{equation}

To this end, we use the doubly robust test proposed by \cite{heterogeneous_ehk} to obtain an asymptotic $p$-value for testing the hypothesis of homogeneous effects (\ref{eq:h0_age}) by using the asymptotic distribution of the statistic
\begin{equation}
\label{eq:test_age}
    T_n = n(C\widehat{\psi})^{T}(C\widehat{\Sigma}C)^{-1}(C\widehat{\psi}),
\end{equation}
where $\widehat{\psi}$ is the vector of the estimated ATT for each age group, and $C$ is a matrix with entries $C_{ij} = 1(i=j) - 1(i=j-1)$, such that hypothesis (\ref{eq:h0_age}) can be written as $C\widehat{\psi} \equiv 0$. Finally, $\widehat{\Sigma}$ is the empirical covariance matrix of $\widehat{\psi}$. 

Under the conditions in \cite{heterogeneous_ehk} we have that 
\begin{equation}
    T_n \rightsquigarrow \chi_{K-1}^2.
\end{equation}
Therefore, in our case, an asymptotic $p$-value for testing (\ref{eq:h0_age}) is given by $\PP(\chi_{7}^2 \geq t_n)$, where $t_n$ is obtained using equation (\ref{eq:test_age}) by replacing the estimated vector of ATT by age. We obtain an observed value $t_n = 14.78$. Thus,
\begin{equation}
    \PP(\chi_{7}^2 \geq 14.78) = 0.0389,
\end{equation}
and we have statistical evidence to reject the null hypothesis of homogeneous ATT by age group.

\end{document}